\documentclass[pra,twocolumn,superscriptaddress,amsmath,amssymb,floatfix]{revtex4}
\usepackage{graphics,pstricks,amsmath}
\usepackage{graphicx}
\usepackage{epsfig}

\begin{document}

\author{Piotr S. \.Zuchowski}
\email{E-mail: Piotr.Zuchowski@durham.ac.uk}
\affiliation{Department of Chemistry, Durham University, South
Road, DH1 3LE, United Kingdom}

\author{Jeremy M. Hutson}
\email{E-mail: J.M.Hutson@durham.ac.uk} \affiliation{Department of
Chemistry, Durham University, South Road, DH1 3LE, United Kingdom}

\title{The prospects for producing ultracold NH$_3$ molecules by
sympathetic cooling: \\ a survey of interaction potentials}

\date{\today}

\begin{abstract}
We investigate the possibility of producing ultracold NH$_3$
molecules by sympathetic cooling in a bath of ultracold atoms.
We consider the interactions of NH$_3$ with alkali-metal and
alkaline-earth atoms, and with Xe, using {\em ab initio}
coupled-cluster calculations. For Rb-NH$_3$ and Xe-NH$_3$ we
develop full potential energy surfaces, while for the other
systems we  characterize the stationary points (global and
local minima and saddle points). We also calculate isotropic
and anisotropic Van der Waals $C_6$ coefficients for all the
systems. The potential energy surfaces for interaction of
NH$_3$ with alkali-metal and alkaline-earth atoms all show deep
potential wells and strong anisotropies. The well depths vary
from 887 cm$^{-1}$ for Mg-NH$_3$ to 5104 cm$^{-1}$ for
Li-NH$_3$. This suggests that all these systems will exhibit
strong inelasticity whenever inelastic collisions are
energetically allowed and that sympathetic cooling will work
only when both the atoms and the molecules are already in their
lowest internal states. Xe-NH$_3$ is more weakly bound and less
anisotropic.
\end{abstract}

\maketitle

\section{Introduction}

There is great interest at present in producing samples of cold
molecules (below 1~K) and ultracold molecules (below 1~mK). Such
molecules have many potential applications. High-precision
measurements on ultracold molecules might be used to measure
quantities of fundamental physics interest, such as the electric
dipole moment of the electron \cite{Hudson:2002} and the
time-dependence of fundamental constants such as the
electron/proton mass ratio \cite{vanVeldhoven:2004}. Ultracold
molecules are a stepping stone to ultracold quantum gases
\cite{Baranov:2002} and might have applications in quantum
information and quantum computing \cite{DeMille:2002}.

There are two basic approaches to producing ultracold
molecules. In {\em direct} methods such as Stark deceleration
\cite{Bethlem:IRPC:2003, Bethlem:2006} and helium buffer-gas
cooling \cite{Weinstein:CaH:1998}, preexisting molecules are
cooled from higher temperatures and trapped in electrostatic or
magnetic traps. In {\em indirect} methods
\cite{Hutson:IRPC:2006}, laser-cooled atoms that are already
ultracold are paired up to form molecules by either
photoassociation \cite{Jones:RMP:2006} or tuning through
magnetic Feshbach resonances \cite{Koehler:RMP:2006}.

Indirect methods have already been used extensively to produce
ultracold molecules at temperatures below 1 $\mu$K. However,
they are limited to molecules formed from atoms that can
themselves be cooled to such temperatures. Direct methods are
far more general than indirect methods, and can in principle be
applied to a very wide range of molecules. However, at present
direct methods are limited to temperatures in the range 10-100
mK, which is outside the ultracold regime. There is much
current research directed at finding second-stage cooling
methods to bridge the gap and eventually allow directly cooled
molecules to reach the region below 1~$\mu$K where quantum
gases can form.

One of the most promising second-stage cooling methods that has
been proposed is {\em sympathetic cooling}. The hope is that, if a
sample of cold molecules in brought into contact with a gas of
ultracold atoms, thermalization will occur and the molecules will
be cooled towards the temperature of the atoms. Sympathetic
cooling has already been used successfully to cool atomic species
such as $^6$Li \cite{Schreck:2001} and $^{41}$K
\cite{Modugno:2001} but has not yet been applied to neutral
molecules.

Sympathetic cooling relies on thermalization occurring before
molecules are lost from the trap. Thermalization requires {\em
elastic} collisions between atoms and molecules to redistribute
translational energy. However, electrostatic and magnetic traps
rely on Stark and Zeeman splittings and trapped atoms and
molecules are not usually in their absolute ground state in the
applied field. Any {\em inelastic} collision that converts
internal energy into translational energy is likely to kick
both colliding species out of the trap. The {\em ratio} of
elastic to inelastic cross sections is thus crucial, and a
commonly stated rule of thumb is that sympathetic cooling will
not work unless elastic cross sections are a factor of 10 to
100 greater than inelastic cross sections for the states
concerned.

Inelastic cross sections for atom-atom collisions are sometimes
strongly suppressed by angular momentum constraints. In
particular, for s-wave collisions (end-over-end angular momentum
$L=0$), pairs of atoms in {\em spin-stretched states} (with the
maximum possible values of the total angular momentum $F$ and its
projection $|M_F|$) can undergo inelastic collisions only by
changing $L$. Cross sections for such processes are very small
because, for atoms in S states, the only interaction that can
change $L$ is the weak dipolar coupling between the electron
spins. However, for molecular collisions the situation is
different: the {\em anisotropy} of the intermolecular potential
can change $L$, and this is usually much stronger than spin-spin
coupling.

It is thus crucial to investigate the anisotropy of the
interaction potential for systems that are candidates for
sympathetic cooling experiments. In experimental terms, the
easiest systems to work with are those in which molecules that
can be cooled by Stark deceleration (such as NH$_3$, OH and NH)
interact with atoms that can be laser-cooled (such as
alkali-metal and alkaline-earth atoms). There has been
extensive work on low-energy collisions of molecules with
helium atoms \cite{Balakrishnan:threshold:1997,
Balakrishnan:1999, Balakrishnan:2000, Bohn:2000,
Balakrishnan:CaH:2003, Krems:henh:2003,
Gonzalez-Martinez:2007}, but relatively little on collisions
with alkali-metal and alkaline-earth atoms. Sold\'{a}n and
Hutson \cite{Soldan:2004} investigated the potential energy
surfaces for Rb + NH and identified deeply bound ion-pair
states as well as weakly bound covalent states. They suggested
that the ion-pair states might hinder sympathetic cooling. Lara
{\em et al.} \cite{Lara:PRL:2006, Lara:PRA:2007} subsequently
calculated full potential energy surfaces for Rb + OH, for both
ion-pair states and covalent states, and used them to
investigate low-energy elastic and inelastic cross sections,
including spin-orbit coupling and nuclear spin splittings. They
found that even for the covalent states the potential energy
surfaces had anisotropies of the order of 500 cm$^{-1}$ and
that this was sufficient to make the inelastic cross sections
larger than inelastic cross sections at temperatures below 10
mK. Tacconi {\em et al.}\ \cite{Tacconi:2007} have recently
carried out analogous calculations on Rb + NH, though without
considering nuclear spin. There has also been a considerable
amount of work on collisions between alkali metal atoms and the
corresponding dimers \cite{Soldan:2002, Quemener:2004,
Cvitas:bosefermi:2005, Cvitas:hetero:2005, Quemener:2005,
Hutson:IRPC:2007}.

One way around the problem of inelastic collisions is to work
with atoms and molecules that are in their absolute ground
state in the trapping field. However, this is quite limiting:
only optical dipole traps and alternating current traps
\cite{vanVeldhoven:2005} can trap such molecules. It is
therefore highly desirable to seek systems in which the
potential energy surface is only weakly anisotropic. The
purpose of the present paper is to survey the possibilities for
collision partners to use in sympathetic cooling of NH$_3$ (or
ND$_3$), which is one of the easiest molecules for Stark
deceleration.

Even if sympathetic cooling proves to be impractical for a
particular system, the combination of laser cooling for atoms
and Stark deceleration for molecules offers opportunities for
studying molecular collisions in a new low-energy regime. For
example, experiments are under way at the University of
Colorado \cite{Lewandowski:priv:2008} to study collisions
between decelerated NH$_3$ molecules and laser-cooled Rb atoms.

There alkali-metal atom + NH$_3$ systems have not been
extensively studied theoretically, though there has been
experimental interest in the spectroscopy of Li-NH$_3$ complex
as a prototype metal atom-Lewis base complex \cite{Wu:01}. Lim
{\em et al.}\ ~\cite{Lim:2007} recently calculated electrical
properties and infrared spectra for complexes of NH$_3$ with
alkali-metal atoms from K to Fr and gave the equilibrium
structures of their global minima. However, to our knowledge,
no complete potential energy surfaces have been published for
any of these systems. The alkaline-earth + NH$_3$ have been
studied even less, and except for an early study of the
Be-NH$_3$ system~\cite{Chalasinski:Be:93} there are no previous
results available.

\section{ { \em Ab initio } methods }

The interaction energy of two monomers $A$ and $B$ is defined as
\begin{equation}
E^{AB}_{\rm int} = E^{AB}_{\rm tot} - E^{A}_{\rm tot} - E^{B}_{\rm
tot}
\end{equation}
where $E^{AB}_{\rm tot}$ is the total energy of the dimer and
$E^{A}_{\rm tot}$ and $E^{B}_{\rm tot}$ are the total energies
of the isolated monomers. Since the interaction energy is
dominated at long range by intermolecular correlation
(dispersion), {\em ab initio} calculations of the interaction
energy must include electronic correlation effects at the
highest possible level \cite{Chalasinski:00} and must be
carried out with large basis sets augmented by diffuse
functions. At present, the coupled-cluster (CC) method with
single, double and noniterative triple excitations (CCSD(T))
provides the best compromise between high accuracy and
computational cost. In the present paper, we carry out
coupled-cluster calculations using the {\sc Molpro} package
\cite{MOLPRO_brief}. All interaction energies are corrected for
basis-set superposition error (BSSE) with the counterpoise
method of Boys and Bernardi \cite{Boys:70}.

Standard coupled-cluster methods are reliable only when the
wavefunction is dominated by a single electronic configuration
This is often an issue for molecular systems with low-lying
excited states. In order to check the reliability of CC
calculations, it is necessary to monitor the norm of $T_1$
operator \cite{Lee:89} (measured by the T1 diagnostic). In the
case of metal-NH$_3$ systems this is relatively large,
especially when the atom approaches the lone pair of the NH$_3$
molecule, but the convergence of the CC equations is fast and
the converged CCSD results are very close to benchmark
multireference configuration interaction (MRCI-SD) calculations
with size-extensivity corrections. Thus we consider the CC
results reliable.

To understand the origin of the intermolecular forces we also
consider the interaction energies obtained at the Hartree-Fock
level, which neglects electron correlation and thus provides
information about the role of dispersion and other correlation
effects. For some systems we also analyze the components of the
intermolecular interactions using symmetry-adapted perturbation
theory \cite{Jeziorski:94} (SAPT). The first-order SAPT
corrections (electrostatic and exchange terms) are computed at
the Hartree-Fock level, while the dispersion energy is
evaluated in the coupled Hartree-Fock approximation
\cite{Zuchowski:03}. These calculations are carried out using
the {\sc SAPT2006}\cite{SAPT:06} program.

We are interested principally in the collisions of cold ammonia
molecules with atoms at energies that are much too low for
vibrational excitation to occur. Such collisions are governed
by an effective potential that is vibrationally averaged over
the ground-state vibrational wavefunction of NH$_3$. For the
present purpose it is adequate to represent this by a potential
calculated with the NH$_3$ molecule frozen at a geometry that
represents the ground state. In the present paper we use a
geometry derived from the high-resolution infrared
spectra~\cite{Benedict:57}: the molecule is taken to have
$C_{3v}$ symmetry with N-H bond lengths of 1.913 $a_0$ and an
H-N-H angle of $106.7^\circ$. Intermolecular geometries are
specified in Jacobi coordinates: $R$ is the distance from the
center of mass of NH$_3$ to the atom, while $\theta$ is the
angle between the intermolecular vector and the $C_3$ axis of
the NH$_3$ molecule (with $\theta=0^\circ$ corresponding to the
atom approaching towards the lone pair of NH$_3$). Finally,
$\chi$ is the dihedral angle between the plane containing the
$C_3$ axis and an NH bond and that containing the $C_3$ axis
and the intermolecular vector.

Table \ref{prop} gives the lowest excitation energies, dipole
polarizabilities and ionization energies of the atoms studied
in this paper. The neutral alkali-metal and alkaline-earth
atoms (denoted below as A and Ae, respectively) have
particularly low excitation energies, resulting from small
separations between energy levels corresponding to $ns$ and $n
p$ or $(n-1) d$ configurations. Since the gap between the
ground and excited states is small, the atoms have very large
polarizabilities. Hence, we expect particularly strong
induction and dispersion interactions. The alkali-metal and
alkaline-earth atoms also have low ionization energies $E_{\rm
i}$. Since the atomic orbital wavefunctions vanish at long
range as $\exp(-E_{\rm i}^{1/2} r)$, the wavefunctions and
densities are very diffuse, and this causes large overlap
between monomers even at relatively large separations. Finally,
because of the low ionization energies, alkali-metal and
alkaline-earth atoms have a strong tendency to form
charge-transfer complexes.

\begin{table*}
\caption{ Properties of alkali-metal, alkaline-earth and Xe
atoms important to interaction potentials. Note that for
alkali-metal atoms the lowest excitation energy corresponds to
$^2S \to\ ^2P_{1/2}$ excitation and for alkaline-earth and Xe
atoms to $^1S \to\ ^3P_0$. The excitation and ionization
energies are take from NIST Handbook of Basic Atomic
Spectroscopic Data \cite{NIST:Phys:2005}.
 } \label{prop}
\begin{ruledtabular}
\begin{tabular}{lrrrrrr}
 Atom &  dipole polarizability & & $C_6$ coefficient   & & lowest excitation energy & ionization energy \\
      &       ($a_0^3$)   & Ref.    & ($E_{\rm h}a_0^6$)  & Ref.  & (cm$^{-1}$)              & (cm$^{-1}$) \\ \hline
Li  & 164   & \onlinecite{Yan:96}  & 1395 &  \onlinecite{Mitroy:03}  &  14904    &   43487  \\
Na  & 162   & \onlinecite{Volz:96}   & 1561 & \onlinecite{Mitroy:03}  &  16956    &   41449  \\
K   & 293   & \onlinecite{Volz:96}   & 3921 & \onlinecite{Pashov:08}  &  12985    &   35010  \\
Rb  & 319   & \onlinecite{Volz:96}   & 4707 & \onlinecite{Marte:02}  &  12578    &   33691  \\
Be  &  37.7 & \onlinecite{Mitroy:03}   & 213  & \onlinecite{Mitroy:03}  &  21978    &   75193  \\
Mg  &  71   & \onlinecite{Mitroy:03}   & 629  & \onlinecite{Mitroy:03}  &  21850    &   61671  \\
Ca  & 159   & \onlinecite{Mitroy:03}   & 2221 & \onlinecite{Porsev:02}  &  15157    &   49305  \\
Sr  & 200   & \onlinecite{Mitroy:03}   & 3250 & \onlinecite{Mitroy:03}  &  14317    &   45932  \\
Xe  & 27.3  & \onlinecite{Lozeille:02}   & 286  & \onlinecite{Tang:03}  &  67068    &   97834  \\
\end{tabular}
\end{ruledtabular}
\end{table*}
%& \onlinecite{PhysRevA.54.2824}
%& \onlinecite{PhysRevA.54.2824}   % RB78 = Rubbmark, J. R. and Borgstrom S. A., Phys Scr 18, 196 (1978)
%& \onlinecite{PhysRevA.54.2824} & % REB95 = Radziemski L. J. , Engleman Jr. R., Brault, J. W. PhysRevA 52,4462 (1995)
%& \onlinecite{PhysRevA.54.2824} & % MZ81 = Martin W.C. Zalubas R. J. Phys. Chem. Ref. Data 10, 153 (1981)
%& \onlinecite{PhysRevA.68.052714} % MZ80 = Martin W.C. Zalubas R. J. Phys. Chem. Ref. Data 9 , 1  (1980)
%& \onlinecite{PhysRevA.68.052714} % SC85 = Sugar J., Corliss, C. J. Phys. Chem. Ref. Data 14 , 2  (1985)
%& \onlinecite{PhysRevA.68.052714} % E99 = Engleman R. ,NIST Atomic Spectra Database
% E99 = Engleman R. ,NIST Atomic Spectra Database
% BVHU01 = Brandi F. , Velchev I. , Horgervorst W. , Ubasch W. , PhysRevA 64,032505
% BVHU01 = Brandi F. , Velchev I. , Horgervorst W. , Ubasch W. , PhysRevA 64,032505
% BBLB98 = Baugh J. F. , Burkhardt C. E., Leventhal J. J., Bergeman T. PhysRevA 58,1585(1998)

The basis sets used in the {\em ab initio} calculations are as
follows. For Be, Li, Mg, Na, Ca  atoms we use all-electron
cc-pVTZ basis sets augmented by even-tempered diffuse
exponents, while for potassium we use the CVTZ basis set of
Feller ~\cite{Feller:95}. For Rb, Sr and Xe we handle only the
outermost electrons explicitly, with the core electrons
represented by effective core potentials (ECPs). For Rb we use
the small-core effective core potential ECP28MWB with a basis
set based on that of Ref.~\onlinecite{Soldan:03}, which was
optimized to recover the static dipole polarizability. We
modified this slightly to account better for intramonomer
electronic correlation effects by removing 0.07 $f$ and adding
0.001049 $s$, 0.0024 $p$, 4.5,0.016667 $d$, 1.9,0.655 $f$ and
0.95,0.3167 $g$ functions. The basis set for Sr is taken from
Ref.~\onlinecite{Lim:06}. For Xe we use the basis set given by
Lozeille {\em et al.}~\cite{Lozeille:02}, which was found to be
excellent for polarizabilities and hyperpolarizabilities. For
each system we added a set of midbond functions with exponents
$sp$: 0.9,0.3,0.1, $df$ 0.6,0.2 to improve the representation
of the dispersion energy in the region of the Van der Waals
minimum.

\section{Results and discussion}

The potential energy surface for an atom-NH$_3$ system is a function
of the intermolecular distance $R$ and two angles $\theta$ and
$\chi$. However, functions of 3 variables are difficult to represent
graphically. It is convenient to represent the $\chi$-dependence in
the form
\begin{equation}
V(R,\theta,\chi)=\sum_{k=0}^\infty V_{3k}(R,\theta) \cos 3k\chi.
\end{equation}
To reduce the computational effort we calculate the interaction
potential only for $\chi=0^\circ$ and $\chi=60^\circ$ and
approximate the leading terms $V_0(R,\theta)$ and $V_3(R,\theta)$ by
sum and difference potentials,
\begin{eqnarray}
V_0(R,\theta) &=& \textstyle{\frac{1}{2}}[V(R,\theta,0)+
V(R,\theta,60^\circ)] \nonumber \\ V_3(R,\theta) &=&
\textstyle{\frac{1}{2}}[V(R,\theta,0)- V(R,\theta,60^\circ)].
\label{V0V3}
\end{eqnarray}
$V_0$ can be viewed as the interaction potential averaged over
$\chi$, while $V_3$ describes the leading anisotropy of the
potential with respect to rotation about the $C_3$ axis of
NH$_3$.

\subsection{ Alkali-metal atom + NH$_3$ interactions }

\begin{figure}
\includegraphics[width=\linewidth]{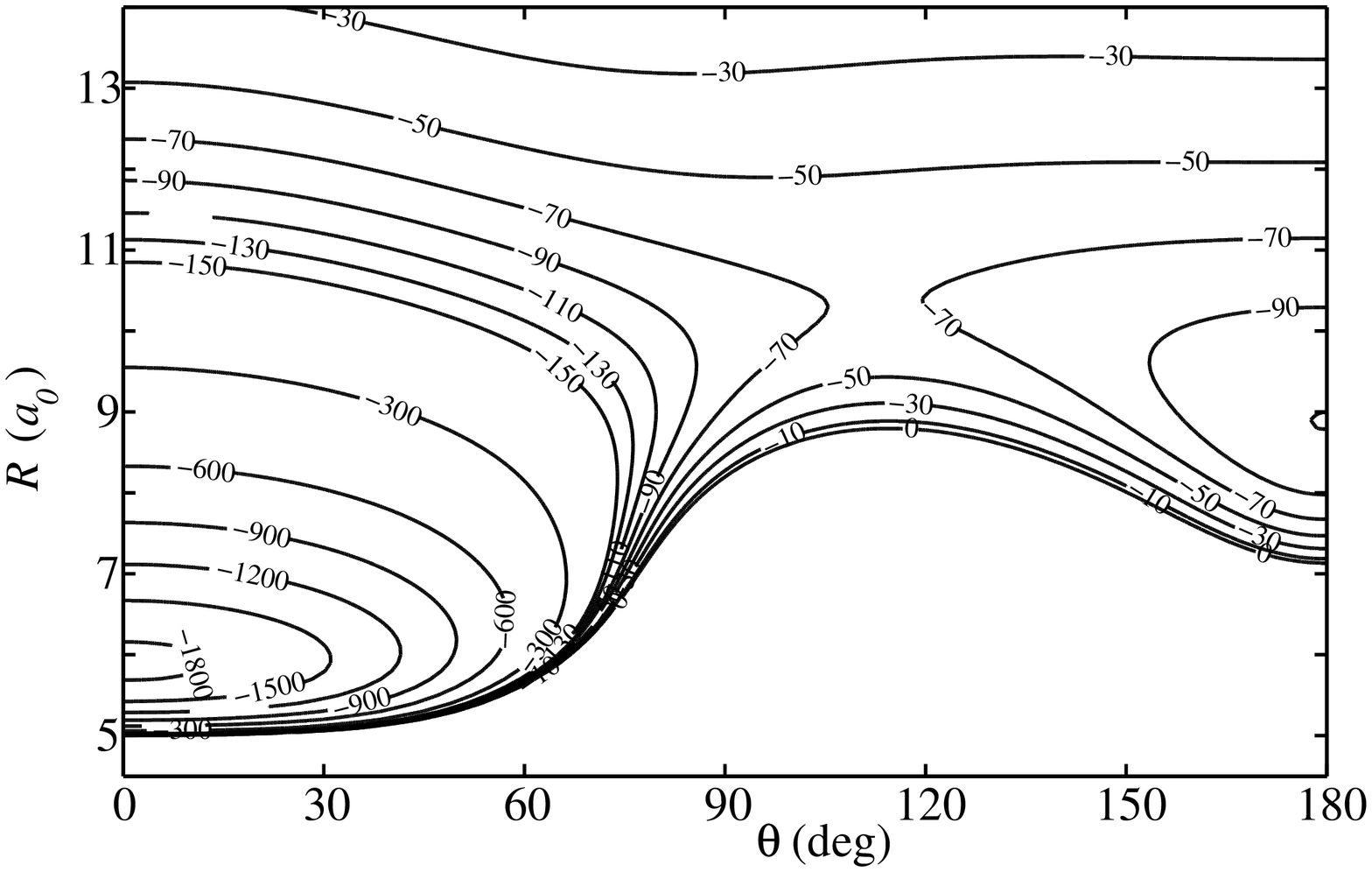}
\includegraphics[width=\linewidth]{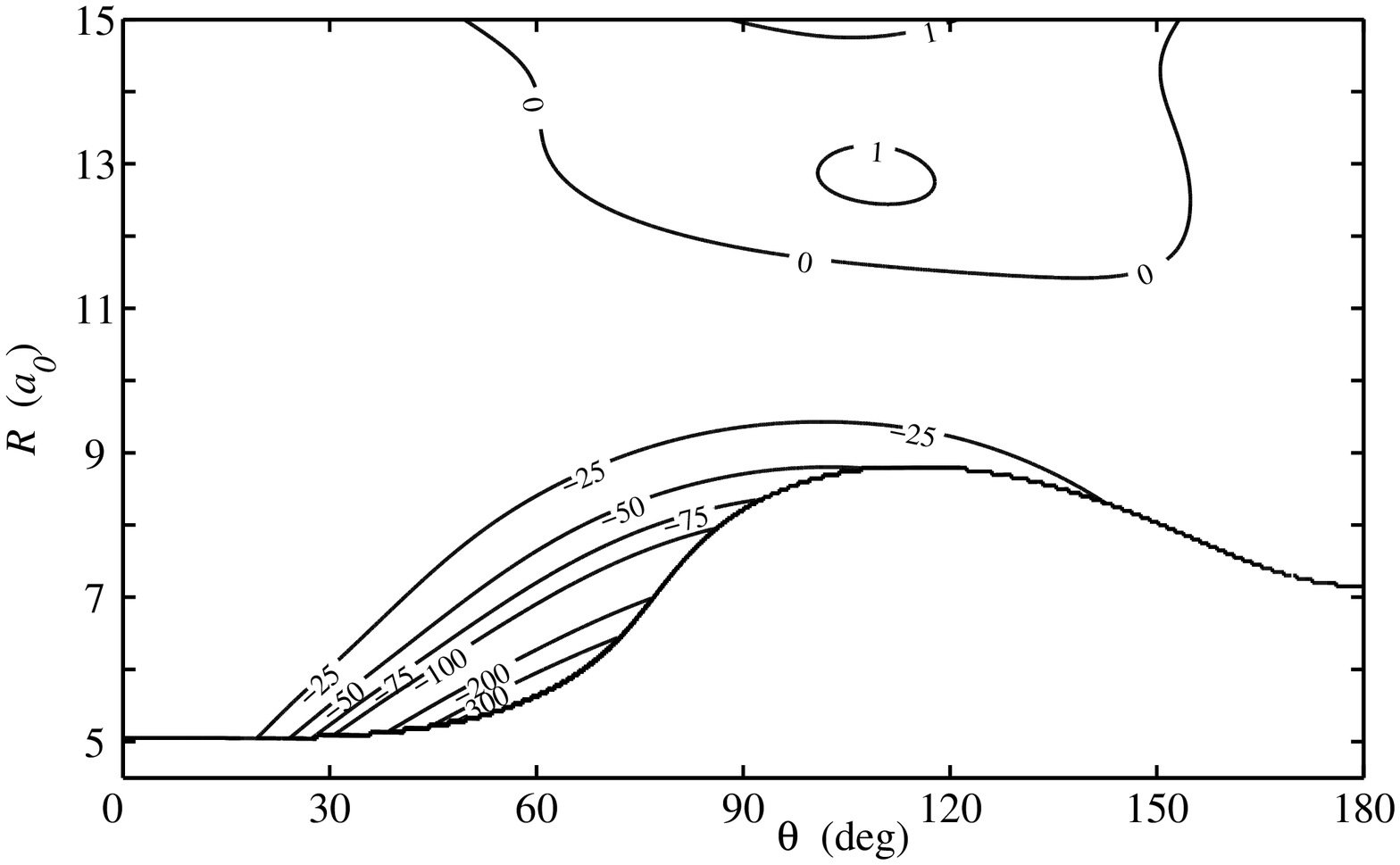}
\caption{ The interaction potential of Rb-NH$_3$ from CCSD(T)
calculations: $V_0(R,\theta)$ component (upper panel) and
$V_3(R,\theta)$ component (lower panel). Contours are labelled in
cm$^{-1}$. To aid visualization, $V_3$ is plotted only in the
energetically accessible region defined by $V_0<0$.} \label{rbnh3}
\end{figure}

The potential energy surface for Rb-NH$_3$ is shown in Figure
\ref{rbnh3}. CCSD(T) calculations were carried out at
$\chi=0^\circ$ and $60^\circ$, at values of $\theta$
corresponding to a 20-point Gauss-Lobatto quadrature. The grid
included $R$ values from 3.5 to $12\ a_0$ with a step of 0.5$\
a_0$, and from 12 to 15 $a_0$ with the step of 1 $a_0$. There
is a deep minimum (1862 cm$^{-1}$) at $R=5.90\ a_0$ and
$\theta=0$, corresponding to approach of Rb towards the NH$_3$
lone pair. The potential is much shallower at other geometries,
with a saddle point near $\theta=110^\circ$ and a shallow
secondary minimum at $\theta=180^\circ$. The anisotropy with
respect to rotation of NH$_3$ about the $C_3$ axis ($\chi$) is
relatively weak, at least in the low-energy classically allowed
region defined by $V_0(R,\theta)<0$.

The overall shape of the other A-NH$_3$ potentials is quite
similar. In each case there is a deep minimum around
$\theta=0^\circ$ and a shallow secondary minimum for
$\theta=180^\circ$. Table \ref{anh3} gives the well depths and
equilibrium distances. For the alkali metals the well depth of
the global minimum decreases down the periodic table, from 5104
cm$^{-1}$ for Li to 1862 cm$^{-1}$ for Rb, and the equilibrium
distance increases from 3.91 $a_0$ for Li to 5.90 $a_0$ for Rb.
The changes in the properties of the shallow secondary minima
are much smaller, with well depths close to 100 cm$^{-1}$ for
all the alkali metals. Our results for the species containing K
and Rb are in good agreement with the CCSD(T) calculations of
Lim {\em et al.}~\cite{Lim:2007}; they obtained slightly
different values of the binding energies of K-NH$_3$ and
Rb-NH$_3$ (2210 cm$^{-1}$ and 1950 cm$^{-1}$, respectively),
but their results are not corrected for BSSE.
 It should also be noted that their binding energies are for
relaxed NH$_3$ geometries.

\begin{table}
\caption{Equilibrium distances and well depths for alkali-metal
atom + NH$_3$ systems from CCSD(T) calculations. } \label{anh3}
\begin{ruledtabular}
\begin{tabular}{lrrrr}
    &  \multicolumn{2}{c}{  $\theta=0^\circ$   }   &  \multicolumn{2}{c}{  $\theta=180^\circ$   }\\

     &   $R_e$\ $(a_0)$    &$D_e$\ (cm$^{-1}$)&   $R_e$\ $(a_0)$   &$D_e$ (cm$^{-1}$) \\ \cline{2-5}
 Li  &   3.91            &      5104        &       7.86       &     104.8     \\
 Na  &   4.73            &      2359        &       8.33       &     98.2      \\
 K   &   5.52            &      2161        &       8.90       &     99.6      \\
 Rb  &   5.90            &      1862        &       8.89       &     110.2     \\
\end{tabular}
\end{ruledtabular}
\end{table}

The deep wells and large anisotropies of the A-NH$_3$
potentials will produce strong coupling between the different
NH$_3$ rotational states during collisions. All these systems
are therefore likely to have large inelastic cross sections. It
is thus unlikely that sympathetic cooling of NH$_3$ with
alkali-metal atoms will be successful unless both the atoms and
the molecules are already are in their lowest internal states.

\subsection{ Alkaline-earth atom + NH$_3$ interactions }

We originally hoped that the potentials for systems containing
alkaline-earth atoms would be more weakly bound and less
anisotropic than for those containing alkali-metal atoms .
However, this proved not to be the case, at least for the
heavier alkaline-earth atoms that are most suitable for laser
cooling. The results for the Ae-NH$_3$ systems are summarized
in Table \ref{aenh3}. The shapes of the potential energy
surfaces are generally similar to those for A-NH$_3$ systems.
For Ca and Sr, the depths of the global minima are 3229 and
3141 cm$^{-1}$ respectively; these are both deeper than for the
corresponding alkali-metal atom. For Mg, however, the well
depth is considerably shallower at only 887.5 cm$^{-1}$. The
minima corresponding to approach at the hydrogen end of NH$_3$
are slightly deeper than for the alkali metals, ranging from
115.7 for Mg to 131.6 cm$^{-1}$ for Sr. On the other hand, the
interaction potential for Be-NH$_3$ resembles those for
Ca-NH$_3$ and Sr-NH$_3$ more than that for Mg-NH$_3$: the
global minimum is 1973 cm$^{-1}$ deep, while the
dispersion-bound minimum is 100.5 cm$^{-1}$ deep. The
equilibrium distance for Be-NH$_3$ at $\theta=0^\circ$ (3.57
$a_0$) is also much shorter than for the other Ae-NH$_3$
systems, and is comparable to that for Li-NH$_3$.

\begin{table}
\caption{Equilibrium distances and well depths for
alkaline-earth atom + NH$_3$ systems from CCSD(T) calculations.
} \label{aenh3}
\begin{ruledtabular}
\begin{tabular}{lrrrr}
     &  \multicolumn{2}{c}{  $\theta=0^\circ$   }   &  \multicolumn{2}{c}{  $\theta=180^\circ$   }  \\
     &   $R_e$\ $(a_0)$    &$D_e$\ (cm$^{-1}$) &   $R_e$\ $(a_0)$    &$D_e$\ (cm$^{-1}$)   \\ \cline{2-5}
 Be  &    3.57           &    1973          &       7.61       &    100.5      \\
 Mg  &    4.83           &     887.5        &       8.20       &    115.7      \\
 Ca  &    4.92           &    3229          &       8.85       &    129.1      \\
 Sr  &    5.22           &    3141          &       9.06       &    131.6      \\
\end{tabular}
\end{ruledtabular}
\end{table}

\subsection { Origin of bonding in metal-atom + NH$_3$ systems}

It is important to understand the large difference between the
metal--lone pair bond energies between Mg and the other Group 1
and 2 atoms considered here. Table \ref{tab_compI} gives the
interaction energies in the global and secondary minima at the
Hartree-Fock, CCSD and CCSD(T) levels for Li-NH$_3$, Mg-NH$_3$
and Ca-NH$_3$. For all these systems the Hartree-Fock
interaction energies are positive for the shallow secondary
minima, indicating that the shallow wells are dominated by
dispersion forces. At the global minima, however, Mg-NH$_3$ is
repulsive at the Hartree-Fock level while the other two systems
are strongly attractive. There is thus strong chemical bonding
in Li-NH$_3$ and Ca-NH$_3$ that is absent in Mg-NH$_3$.

\begin{table}
\caption{The interaction energies (in cm$^{-1}$) for Li-NH$_3$,
Ca-NH$_3$ and Mg-NH$_3$ at different levels of electronic
correlation, for geometries corresponding to the global and
secondary minima.} \label{tab_compI}
\begin{ruledtabular}
\begin{tabular}{lrrr}
     &  \multicolumn{3}{c}{  GM   }       \\
     &    HF      &   CCSD    &   CCSD(T) \\ \hline
 Li  &  -4405     &   -5022   &  -5104    \\
 Ca  & -2152      &   -2937   &  -3229    \\
 Mg  &  260       &   -590    &   -888    \\ \hline
     &  \multicolumn{3}{c}{  LM  }        \\
     &    HF      &   CCSD    &   CCSD(T) \\ \hline
 Li  &  248       &  -54      &  -105     \\
 Ca  &  244       &  -47      &  -129     \\
 Mg  &  155       &  -54      &  -116     \\
\end{tabular}
\end{ruledtabular}
\end{table}

\begin{figure}
\includegraphics[width=\linewidth]{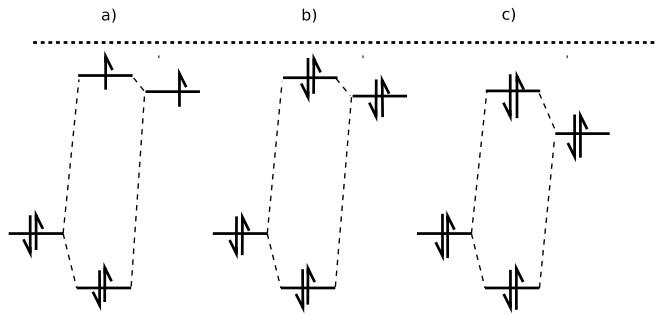}
\caption{ The pattern of molecular orbitals for a) Li-NH$_3$, b)
Ca-NH$_3$, c) Mg-NH$_3$ near their global minima. The
HOMOs of NH$_3$ and of the metal atoms form bonding and
antibonding orbitals. Note the small change in the HOMO energy for
the Li-NH$_3$ and Ca-NH$_3$ systems and the much larger change for
Mg-NH$_3$. } \label{molevels}
\end{figure}

The qualitative differences between Mg and the other atoms can
be understood if we consider how the energy of the highest
occupied molecular orbital (HOMO) differs for the different
atom-NH$_3$ systems. Fig.\ \ref{molevels} shows the two highest
occupied molecular orbitals of each system. As we separate the
monomers to infinity, these two orbitals became HOMOs of the
atom and the NH$_3$ molecule. For any alkali-metal atom, the
strong A-NH$_3$ bond can be explained in terms of LCAO-MO
theory as a chemical bond of order one half, since we have a
doubly occupied bonding orbital and a singly occupied
antibonding orbital [see Fig.\ \ref{molevels} a)]. However,
this explanation does not apply to the alkaline-earth atoms,
where the antibonding orbital is doubly occupied. The net
bonding in Ca-NH$_3$ arises because the bonding orbital is
shifted down in energy considerably more than the antibonding
orbital is shifted up. Conversely, in Mg-NH$_3$, the
contributions from the bonding and antibonding orbitals are
closely balanced. The difference can probably be attributed to
the participation of $np$ orbitals; as shown in Table
\ref{prop}, the $S\to P$ splitting is considerably smaller in
Ca than in Mg. Thus Mg-NH$_3$ is bound mainly by dispersion
forces whereas Ca-NH$_3$ has substantial chemical bonding.

%
%Table \ref{tab_compI} also shows large differences between the
%CCSD and CCSD(T) interaction energies for the shallow secondary
%minima. These differences arise from three-body electron
%interactions which cannot be properly described at the CCSD
%(singles and doubles) level. On the other hand, the good
%performance of CCSD near the global minimum again suggests the
%dominance of the chemical bonding, polarization and induction
%effects over dispersion, except for Mg where the difference
%between the CCSD and CCSD(T) interaction energies is significantly
%larger (45\%) than for Ca-NH$_3$ (9\%) and Li-NH$_3$ (1\%).

Different considerations apply to the Be atom, which is a
notoriously difficult case for electronic structure theory
\cite{Patkowski:07}. Although the potential energy surfaces are
qualitatively similar for the Be-NH$_3$, Ca-NH$_3$ and
Sr-NH$_3$ systems at the CCSD(T) level, the origin of the
strong bonding is probably different in Be-NH$_3$. In this case
the Hartree-Fock and CCSD potential energy curves for
$\theta=0^\circ$ show a double-minimum structure, with a
shallow long-range minimum separated from the global minimum by
a barrier. This suggests a sudden change in chemical character
as the Be atom approaches N. At the Hartree-Fock level the
maximum has an energy of 730 cm$^{-1}$ at $R=4.88$ $a_0$. The
long-range minimum at the Hartree-Fock level is 18.4 cm$^{-1}$
deep at $R=9.02$ $a_0$, while at the CCSD level it is 138
cm$^{-1}$ deep at $R=6.5$ $a_0$. Despite this peculiar
behavior, the CC calculations showed no convergence problems or
unusually large T1 diagnostics. However, our results for
Be-NH$_3$ disagree with those of Cha\l asi\'nski and coworkers
\cite{Chalasinski:Be:93}, who carried out fourth-order
M\/oller-Plesset (MP4) calculations and found a global minimum
that corresponds to the outer minimum on the CCSD potential
energy curve. They did not find the inner minimum, which turned
out to be the global minimum in our calculations.

%The first excited state of $A'$ symmetry in Be-NH$_3$ was found
%to be high (over 26000 cm$^{-1}$) above the ground state in MRCI
%calculations.

\begin{figure}
\includegraphics[width=\linewidth]{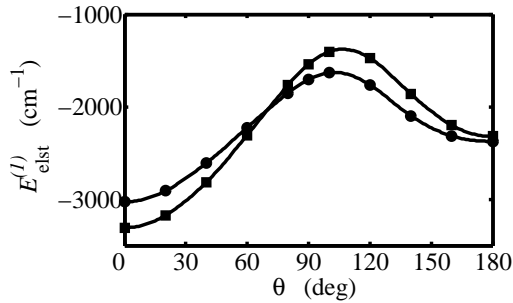}
\caption{Electrostatic energy of Na (squares) and Mg (circles)
atoms interacting with NH$_3$ as a function of $\theta$ for
$R=6 a_0$. The energy is averaged over $\chi$. } \label{epol}
\end{figure}

\begin{figure}
\includegraphics[width=\linewidth]{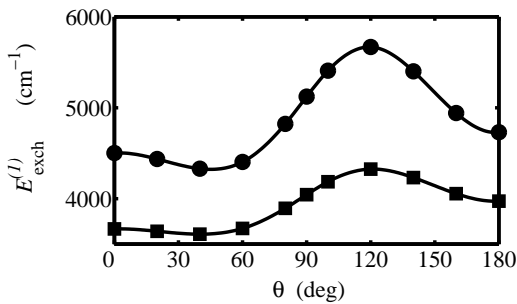}
\caption{First-order exchange energy of Na (squares) and Mg
(circles) atoms interacting with NH$_3$ as a function of
$\theta$ for $R=6 a_0$. The energy is averaged over $\chi$. }
\label{eexch}
\end{figure}

\begin{figure}
\includegraphics[width=\linewidth]{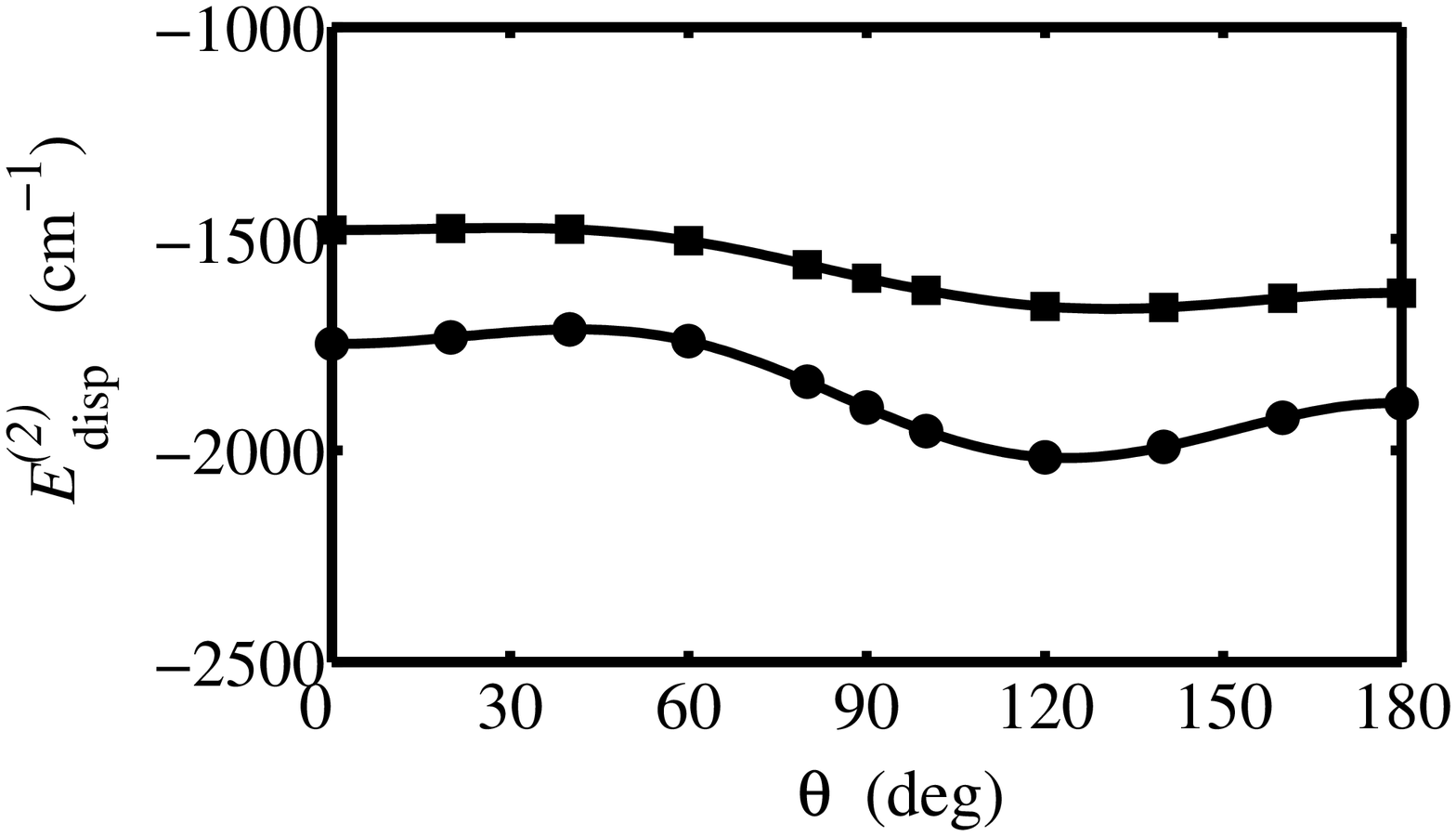}
\caption{Dispersion energy of Na (squares) and Mg (circles)
atoms interacting with NH$_3$ as a function of $\theta$ for
$R=6 a_0$. The energy is averaged over $\chi$. } \label{edisp}
\end{figure}

As mentioned before, the feature of the potential energy
surfaces that is important for elastic/inelastic collision
ratios is the anisotropy. In order to understand the origin of
the anisotropies better, we carried out additional calculations
based on symmetry-adapted perturbation theory (SAPT). Figures
\ref{epol}, \ref{eexch}, and \ref{edisp} show the
electrostatic, first-order exchange and dispersion components
of the interaction energy $V_0$ for Na-NH$_3$ and Mg-NH$_3$,
averaged over $\chi$ as in Eq.\ (\ref{V0V3}). The calculations
were performed at a fixed $R$ value of 6 $a_0$, which is in an
attractive region for $\theta=0^\circ$ and a repulsive region
for $\theta=180^\circ$. Figs.\ \ref{epol}, \ref{eexch} and
\ref{edisp} show clearly that it is the first-order interaction
energy that is responsible for most of the anisotropy in the
valence overlap region. This is caused by a very large
difference between the electrostatic attraction near the
lone-pair site and near hydrogen sites (see Fig.\ \ref{epol}).
This difference is significantly larger than that in the
exchange energy. The anisotropy of the dispersion interaction
(plotted in Fig.\ \ref{edisp}), is even weaker.

%Figs.\ \ref{epol} and \ref{eexch} show very
%strong anisotropy in the first-order interaction energies with
%respect to $\theta$. Note that the difference between the exchange
%energy for $\theta=0^\circ$ and $\theta=180^\circ$ is smaller than
%for the electrostatic component.
%The exchange energy is slightly
%less repulsive for $\theta=0^\circ$ than for $\theta=180^\circ$ but
%the electrostatic energy is much more attractive near
%$\theta=0^\circ$, and the electrostatic attraction
%between the atom and the lone pair is stronger for Na than for Mg.
%This can be explained by the larger extent of the electronic cloud
%of Na atom, and larger charge overlap between Na and the lone pair.
%On the other hand, the Pauli repulsion, caused mostly by the
%first-order exchange energy, is much stronger for Mg. Since Mg is a
%closed-shell atom, the exchange is stronger than for Na.

\begin{figure}
\includegraphics[width=\linewidth]{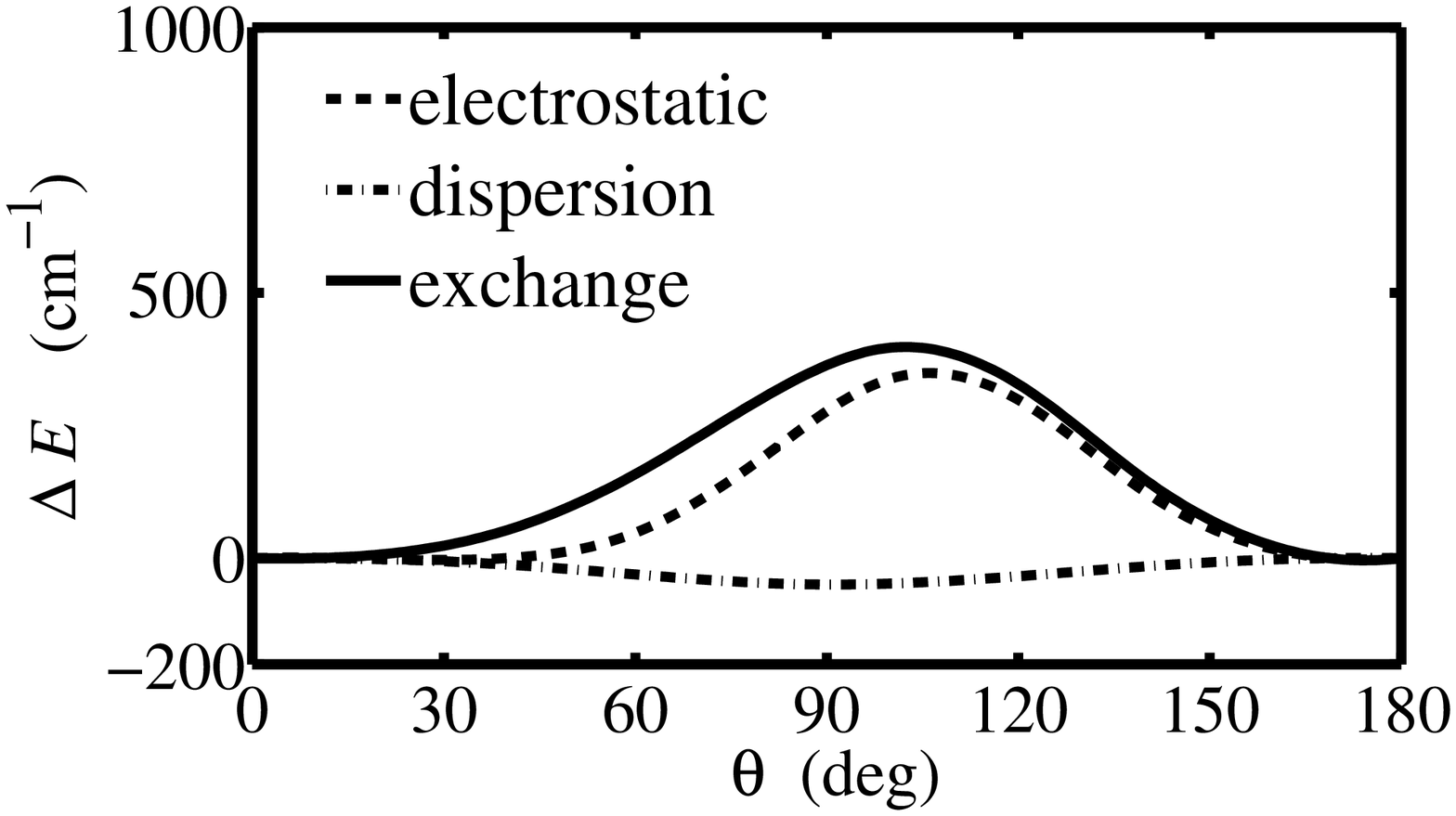}
\includegraphics[width=\linewidth]{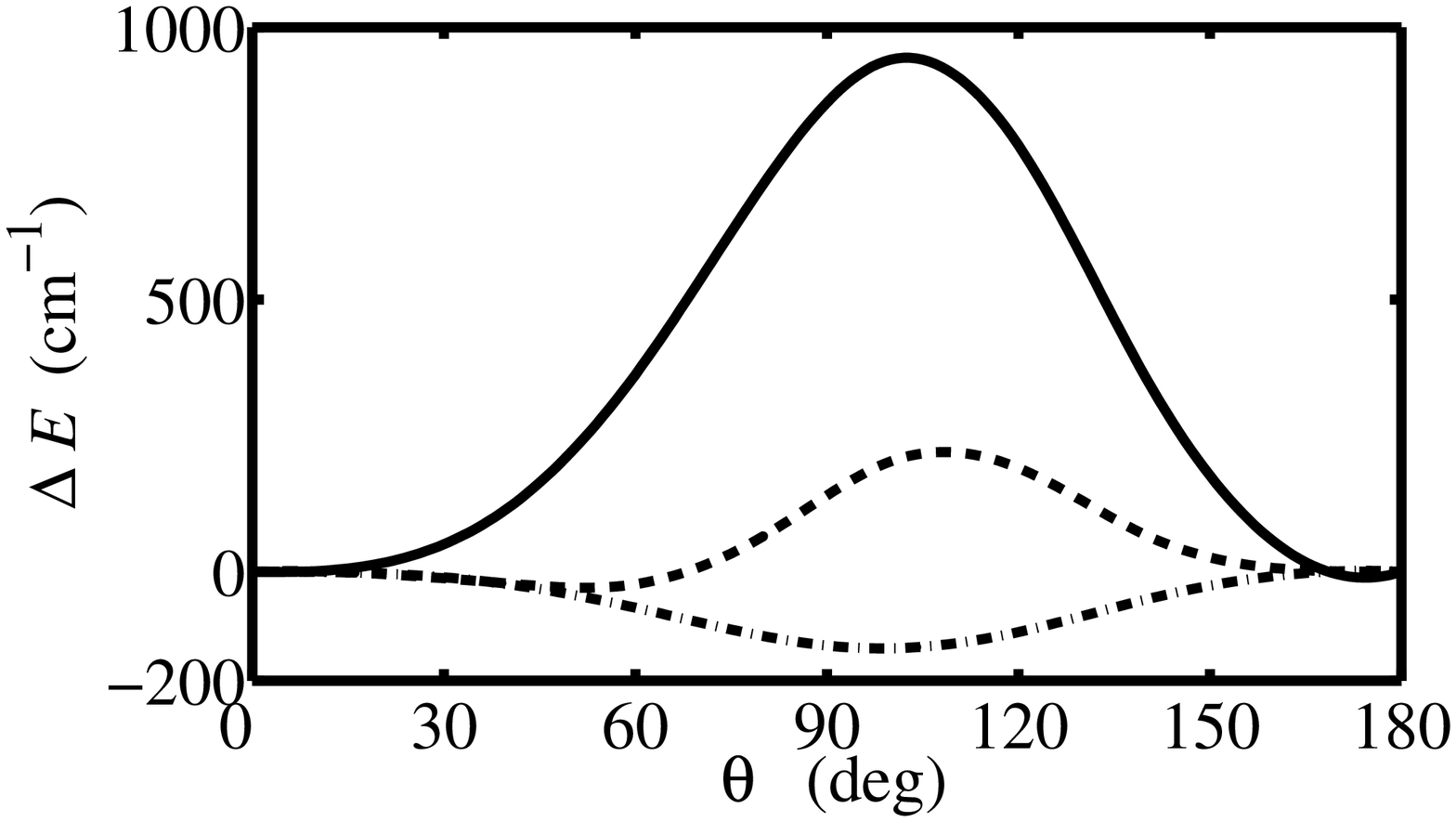}
\caption{Anisotropy of the electrostatic, exchange and
dispersion contributions to the interaction energy, with
respect to rotation about the $C_3$ axis of NH$_3$, as a
function of $\theta$, for $R=6\ a_0$, for Na-NH$_3$ (upper
panel) and Mg-NH$_3$ (lower panel). } \label{aniso}
\end{figure}

The anisotropy $V_3$ of all three components of the interaction
energy with respect to $\chi$ is shown for Na-NH$_3$ and
Mg-NH$_3$ in Fig.\ \ref{aniso}. The exchange energy is very
strongly anisotropic, especially for Mg-NH$_3$. The large
difference in exchange energy between Mg-NH$_3$ and Na-NH$_3$
can be explained by the closed-shell character of the Mg atom
and the much stronger Pauli repulsion between hydrogens of
NH$_3$ and Mg. The electrostatic and dispersion contributions
to $V_3$ are much more similar for the two systems.

\subsection{Xe + NH$_3$ interaction}

All the metal-NH$_3$ systems investigated above have
disappointingly large anisotropies. It is likely that all of
them will exhibit large inelastic cross sections for any
initial state where inelasticity is possible. We therefore
decided to consider other possible collision partners for
sympathetic cooling of NH$_3$. Barker \cite{Barker:priv:2008}
has suggested an experiment in which Xe is first laser-cooled
in its metastable $^3P_2$ state and then transferred to its
ground $^1S_0$ state by laser excitation followed by
spontaneous emission. Since ground-state Xe has a fairly large
dipole polarizability, it can be held in an optical dipole trap
and might be used for sympathetic cooling. In this subsection
we investigate the Xe-NH$_3$ interaction in order to evaluate
its potential in this respect.

Interactions between noble gases and ammonia have been studied
extensively. The interaction between He and NH$_3$ is important
in understanding the spectroscopy of NH$_3$ molecules in helium
nanodroplets~\cite{Behrens:98}. The most recent {\em ab initio}
calculations of Hodges and Wheatley
\cite{CPL.313.313.1999,Hodges:01} gave a global minimum about
33 cm$^{-1}$ deep at $R=6$ $a_0$, $\theta=90^\circ$ and
$\chi=60^\circ$. The interaction of Ar with NH$_3$ has been
studied even more extensively, both experimentally
\cite{JCP.101.146.1994} and by {\em ab initio} methods
\cite{JCP.91.7809.1989, JCP.101.1129.1994}. Inversion of
vibration-rotation-tunnelling spectra~\cite{JCP.101.146.1994}
gave a minimum 147 cm$^{-1}$ deep at $R = 6.5\ a_0$,
$\theta=97^\circ$ and $\chi=60^\circ$, while the {\em ab
initio} MP4 (fourth-order M\o ller-Plesset) calculations of Tao
and Klemperer \cite{JCP.101.1129.1994} gave a global minimum
130 cm$^{-1}$ deep at  $R=6.85\ a_0$, $\theta=90^\circ$ and
$\chi=60^\circ$. The Ne-NH$_3$ system was investigated through
MP4 calculations by van Wijngaarden and J\"ager
\cite{Wijngaarden:01}, who obtained a global minimum 63
cm$^{-1}$ deep at $R=6.1\ a_0$, $\theta=90^\circ$ and
$\chi=60^\circ$. For Kr-NH$_3$, Cha\l asi\'nski {\em et
al.}~\cite{Chalasinski:92} obtained a global minimum 108
cm$^{-1}$ deep at $R=7.2\ a_0$, $\theta=100^\circ$ and
$\chi=60^\circ$. However their results were based on
calculations at the MP2 level and may not reproduce the
dispersion energy accurately.

\begin{figure}
\includegraphics[width=\linewidth]{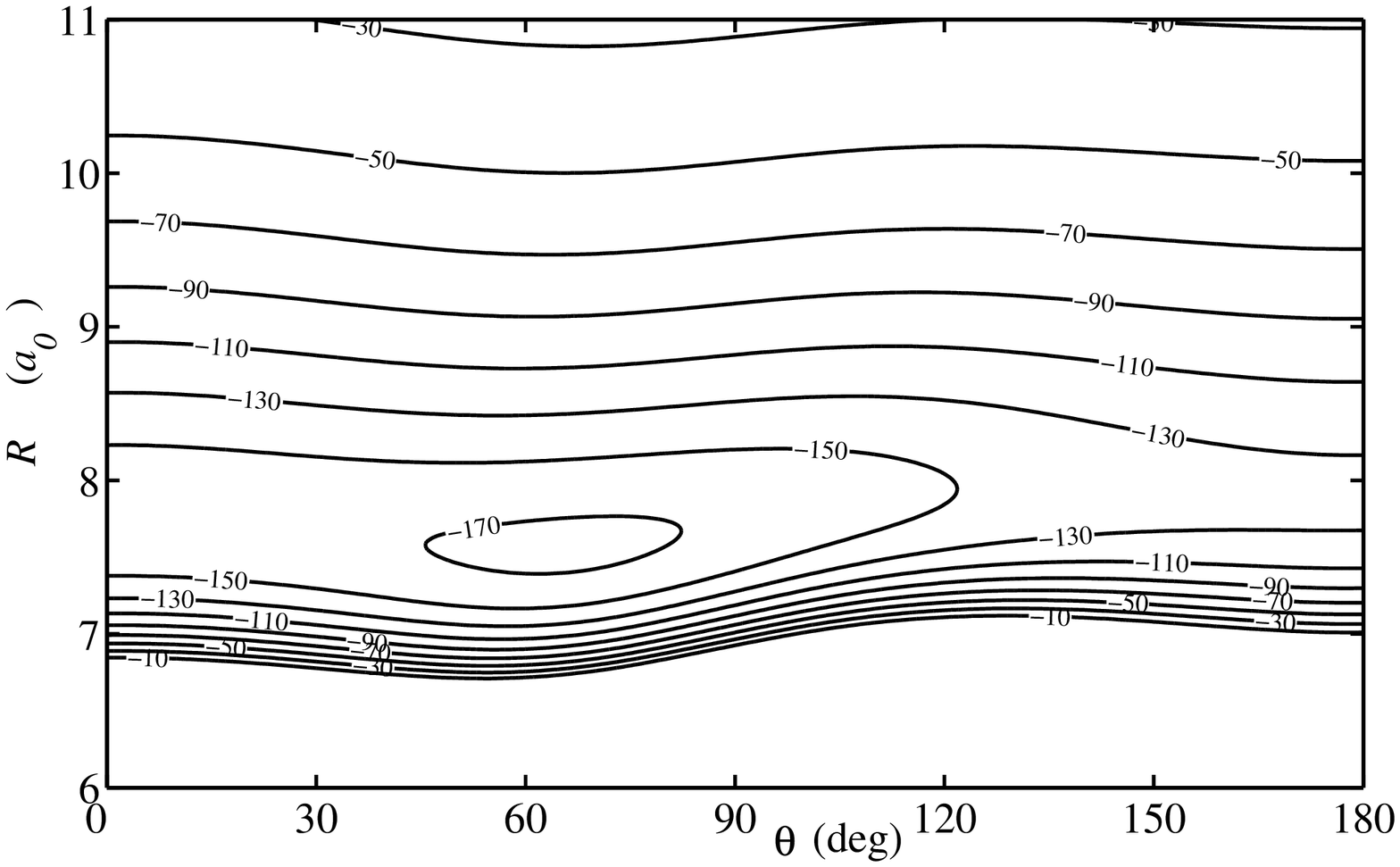}
\includegraphics[width=\linewidth]{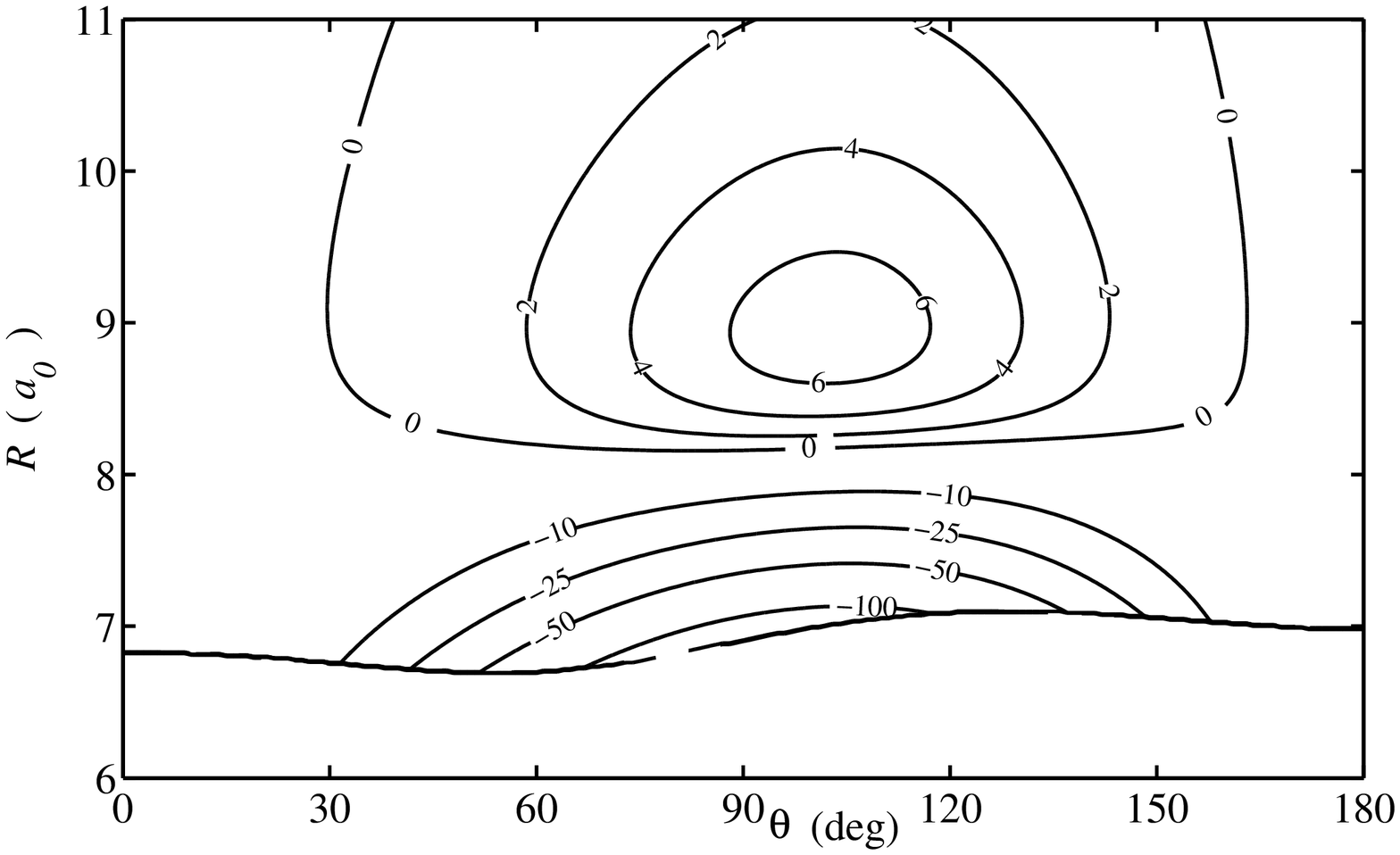}
\caption{The interaction potential of Xe-NH$_3$ from CCSD(T)
calculations: $V_0(R,\theta)$ component (upper panel) and
$V_3(R,\theta)$ component (lower panel). Contours are labelled in
cm$^{-1}$. To aid visualization, $V_3$ is plotted only in the
energetically accessible region defined by $V_0<0$.} \label{xenh3}
\end{figure}

Fig.\ \ref{xenh3} shows the interaction potential for Xe-NH$_3$
from our CCSD(T) calculations. The potential energy surface
differs qualitatively from those for metal-NH$_3$ potentials
studied in the previous subsection, and behaves analogously to
those for other Rg-NH$_3$ systems. The $V_0$ surface for
Xe-NH$_3$ has only one minimum, 173.5 cm$^{-1}$ deep, at
$R=7.65\ a_0$ and $\theta=66^\circ$. The global minimum for the
non-expanded surface is 196.8 cm $^{-1}$ deep, at $R=7.35\
a_0$, $\theta=81^\circ$ and $\chi=60^\circ$. There are saddle
points at both $C_{3v}$ geometries. For $\theta=0$ the saddle
point is 166.2 cm$^{-1}$ deep at $R=7.73\ a_0$, while for
$\theta=180^\circ$ the saddle point is 134.1 cm$^{-1}$ deep at
$R=7.93\ a_0$. The major binding arises from the dispersion
energy, and at the Hartree-Fock level we observe only a small
attraction (a few cm$^{-1}$) at large distances, due to weak
induction forces which behave asymptotically as $-C_6 R^{-6}$.
Near the Van der Waals minimum predicted by CCSD(T), the SCF
energy is repulsive.

The $V_0$ surface for Xe-NH$_3$ system thus has an anisotropy
of only about 60 cm$^{-1}$ between the potential minimum and
the higher of the two saddle points. This is considerably
smaller than for Rb-OH or any of the metal-NH$_3$ systems
studied here, but still substantial compared to the rotational
constant of NH$_3$, $b=6.35$ cm$^{-1}$ for rotation about an
axis perpendicular to $C_3$.

\subsection{Long-range forces}

Long-range forces are very important in cold and ultracold
collisions. We therefore carried out separate calculations of
the Van der Waals coefficients for the interactions. The
isotropic $C_{6,0}$ and anisotropic $C_{6,2}$ dispersion
coefficients for the interaction of atom $A$ and symmetric top
molecule $B$ may be written in terms of the dynamic
polarizabilities of the monomers, evaluated at imaginary
frequencies,
\begin{eqnarray}
C^{\rm disp}_{6,0} &=& \frac{3}{\pi} \int_0^{+\infty} \alpha_A ( i u) \bar{ \alpha}_B(i u) {\rm d } u; \nonumber\\
C^{\rm disp}_{6,2} &=& \frac{1}{\pi} \int_0^{+\infty} \alpha_A ( i u) \Delta{ \alpha}_B(i u) {\rm d } u,
\label{Intgr}
\end{eqnarray}
where $\bar{ \alpha}=\frac{1}{3}(2\alpha_{xx}+\alpha_{zz})$ is
the isotropic polarizability and
$\Delta\alpha=\alpha_{zz}-\alpha_{xx}$ is the polarizability
anisotropy. The induction contributions to the Van der Waals
coefficients are
\begin{equation}
C_{6,0}^{\rm ind} =C_{6,2}^{\rm ind} = \alpha_A \mu^2,
\end{equation}
where the dipole moment $\mu$ is 0.579 $e a_0$ for NH$_3$
\cite{Schmuttenmaer:91}.

The integrals in Eqs.\ \ref{Intgr} were evaluated using the
method given by Amos {\em et al.} ~\cite{Amos:84}. The dynamic
polarizabilities of NH$_3$ were obtained using coupled
Kohn-Sham theory with the asymptotically  corrected PBE0
functional \cite{Adamo:99a} and d-aug-cc-pVTZ basis sets. To
get the dynamic polarizabilities for the alkali-metal atoms, we
adjusted the fraction of exchange, exact exchange and
correlation fraction in the PBE0 functional in such a way as to
recover the atom-atom $C_6$ coefficients (see Table
\ref{prop}). Our coupled Kohn-Sham program does not allow us to
use core potentials to calculate dynamic polarizabilities. For
Rb we therefore performed all-electron calculations with the
pVTZ basis set of Sadlej \cite{Sadlej} combined with the
Douglas-Kroll approximation \cite{Douglas:74}. The dynamic
polarizabilities obtained in this way were tested by comparing
$C_6$ coefficients for A-Ar and A-Xe systems with those
obtained by Mitroy and Zhang ~\cite{Mitroy:07}. The maximum
error was found to be +6.3\% (for Na-Xe) while the average
error is less than +3\%. For alkaline-earth and Xe atoms the
frequency-dependent dipole polarizabilities were obtained from
time-independent coupled-cluster linear response functions
\cite{Moszynski:05,Korona:06}.

The resulting $C_6$ coefficients are shown in Table
\ref{vdWcoeff}. For the alkali-metal and alkaline-earth atoms,
the isotropic dispersion coefficients $C^{\rm disp}_{6,0}$ are
fairly large because of the large atomic polarizabilities. The
anisotropies in the dispersion coefficients are much smaller,
because of the small polarizability anisotropy of NH$_3$ (2.1
$a_0^3$) compared to its isotropic polarizability (14.6
$a_0^3$). The induction Van der Waals coefficients are large,
and account for 10-25\% of the total $C_{6,0}$ and 70-90\% of
the total $C_{6,2}$. It may be noted that $C^{\rm disp}_{6,0}$
for Rb-NH$_3$ is somewhat larger than $C^{\rm disp}_{6,00}$ for
Rb-OH \cite{Lara:PRA:2007}. As one might expect, the Xe-NH$_3$
long-range interaction has slightly different character from
the A- and Ae-NH$_3$ systems. The $C^{\rm disp}_{6,0}$
coefficient is still large, but the total anisotropy (in
particular the dispersion anisotropy) is much smaller.

\begin{table}
\caption{Van der Waals dispersion and induction coefficients
for A-NH$_3$ and Ae-NH$_3$ systems. All values are in atomic
units, $E_{\rm h}a_0^6$. } \label{vdWcoeff}
\begin{ruledtabular}
\begin{tabular}{lrrrr}
     &  $C^{\rm disp}_{6,0}$    &   $C^{\rm disp}_{6,2}$  &  $C^{\rm ind}_{6,0}=C^{\rm ind}_{6,2}$       \\ \cline{2-4}
 Li  &     224           &     7.2          &    55.0       \\
 Na  &     258           &     7.4          &    54.3       \\
 K   &     378           &     11.6         &    98.2       \\
 Rb  &     416           &     12.5         &    106.9      \\
 Be  &     121           &      2.3          &    12.7       \\
 Mg  &     200           &      4.4          &    24.0       \\
 Ca  &     342           &      8.4         &     53.4       \\
 Sr  &     413           &     10.2         &     67.4       \\
 Xe  &     161           &     0.94         &      9.1       \\
\end{tabular}
\end{ruledtabular}
\end{table}

\section{Conclusions}

We have investigated the intermolecular potential energy
surfaces for interaction of NH$_3$ with several different atoms
that might be used for sympathetic cooling. For interaction
with all the alkali-metal and alkaline-earth atoms, we found
deep minima and strong anisotropies. The shallowest potential
is for Mg-NH$_3$, but even there the anisotropy in the well
depth is close to 800 cm$^{-1}$. This is likely to cause strong
inelastic collisions for all initial states for which they are
energetically allowed. Accordingly, we consider that none of
the alkali metals and alkaline earths are good prospects for
sympathetic cooling of NH$_3$ unless both the atoms and the
molecules are in their lowest states in the trapping field.
This suggests that sympathetic cooling would need to be carried
out in either optical or alternating current traps.

A somewhat more promising system for sympathetic cooling is
Xe-NH$_3$, for which the global minimum is calculated to be
196.8 cm$^{-1}$ deep at an off-axis geometry. The Xe-NH$_3$
system is relatively weakly anisotropic, with the saddle points
for C$_{3v}$ geometries only 30.6 and 62.7 cm$^{-1}$ higher
than the global minimum. In future work we will use the
interaction potential to calculate low-energy elastic and
inelastic cross sections, in order to predict whether
sympathetic cooling of NH$_3$ by Xe is likely to be feasible.

Even if sympathetic cooling proves to be impossible for these
systems, there is much to be learnt from collisions between
velocity-controlled beams of molecules and laser-cooled atoms.
There are opportunities to explore low-energy inelastic
processes in novel collisional regimes and to probe scattering
resonances in unprecedented detail. We therefore intend to use
the potential energy surfaces developed here to carry out
inelastic collision calculations to explore these effects and
assist in the interpretation of collision experiments.

\bibliography{paper_NH3}
\clearpage

\end{document}